%% file: main.tex
\documentclass[letterpaper, 10 pt, conference]{ieeeconf} 

\IEEEoverridecommandlockouts
\usepackage{amsmath}
\usepackage{amsfonts}
\usepackage{graphicx}
\usepackage{booktabs}
\usepackage{multirow}
\usepackage{url}
\usepackage{listings}
\usepackage{hyperref}
\usepackage{textcomp}
\usepackage{gensymb}

\usepackage{soul}

\title{\LARGE \bf
HackCar: a test platform for attacks and defenses on a cost-contained automotive architecture}

\author{Dario Stabili$^{1}$, Filip Valgimigli$^{2}$, Edoardo Torrini$^{2}$ and Mirco Marchetti$^{2}$
\thanks{$^{1}$Dario Stabili is with the Department of Computer Science and Engineering,
        Alma Mater Studiorum - University of Bologna
        {\tt\small dario.stabili@unibo.it}}%
\thanks{$^{2}$Filip Valgimigli, Edoardo Torrini and Mirco Marchetti are with the Department of Engineering ``Enzo Ferrari'',
        University of Modena and Reggio Emilia
        {\tt\small filip.valgimigli@unimore.it, 123456@studenti.unimore.it, mirco.marchetti@unimore.it}}%
}

\begin{document}
\maketitle
\thispagestyle{empty}
\pagestyle{empty}

\input{sections/abstract}
\input{sections/introduction}
\input{sections/implementation}
\input{sections/experimental_evaluation}
\input{sections/conclusions}

\section*{Acknowledgments}
This work was partially supported by projects SERICS (PE00000014) under the MUR National Recovery and Resilience Plan funded by the European Union - NextGenerationEU and FuSeCar funded by the MIUR Progetti di Ricerca di Rilevante Interesse Nazionale (PRIN) Bando 2022 - grant 2022W3EPEP.

\small{
\bibliographystyle{IEEEtran}
\bibliography{sections/bibliography}{}
}

\end{document}

%% file: sections/abstract.tex
\begin{abstract}
In this paper, we introduce the design of \textit{HackCar}, a testing platform for replicating attacks and defenses on a generic automotive system without requiring access to a complete vehicle. This platform empowers security researchers to illustrate the consequences of attacks targeting an automotive system on a realistic platform, facilitating the development and testing of security countermeasures against both existing and novel attacks.
The \textit{HackCar} platform is built upon an $F1-10^{th}$ model, to which various automotive-grade microcontrollers are connected through automotive communication protocols. This solution is crafted to be entirely modular, allowing for the creation of diverse test scenarios. Researchers and practitioners can thus develop innovative security solutions while adhering to the constraints of automotive-grade microcontrollers.
We showcase our design by comparing it with a real, licensed, and unmodified vehicle. Additionally, we analyze the behavior of the \textit{HackCar} in both an attack-free scenario and a scenario where an attack on in-vehicle communication is deployed.
\end{abstract}

%% file: sections/introduction.tex
\section{Introduction}
\label{s:introduction}
The integration of microcontrollers into modern automobiles has enabled car manufacturers to introduce innovative features aimed at enhancing safety and driving comfort. These functionalities are executed on microcontrollers, commonly known as Electronic Control Units (ECUs), which interface with the mechanical components of the vehicle and communicate through various networks. Among these networks, the most widely adopted one is the Controller Area Network (CAN), developed by Bosch GmbH in the early 90s~\cite{BOSCHCanSpecV2}. While CAN has proven effective in ensuring robust communication in the automotive industry, it lacks essential security assurances crucial for modern applications.

There is a widespread acknowledgment that vehicles are susceptible to hijacking through the injection of maliciously fabricated messages on the CAN bus, a vulnerability demonstrated by security researchers through technical reports and white papers~\cite{MillerValasekAdventures,KeenSecurity}. These attacks exploit the \textit{drive-by-wire} capabilities of modern vehicles, allowing control of the driving system through messages transmitted over the CAN bus. For example, speed control, activated to maintain a constant vehicle speed for reduced fuel consumption and emissions, serves as a practical example of a well-known and widely used \emph{drive-by-wire} capability. Despite these systems being developed for safety and comfort, their deployment fundamentals open avenues for targeted attacks, posing risks to the safety of individuals both inside and outside the vehicle.

With growing concerns about vehicle security, numerous researchers and car manufacturers are actively engaged in fortifying the CAN network by implementing diverse security measures, including intrusion detection systems~\cite{Kneib2018Scission, Stabili2022DAGA}, encryption~\cite{Groza2012LibraCAN}, and authentication~\cite{Kurachi2016CaCAN}. Following the initial public demonstration of a remote attack on a modern connected vehicle~\cite{MillerValasekBlackHatPaper}, cybersecurity researchers have proposed various detection algorithms tailored for in-vehicle communication networks to identify cyber-attacks~\cite{Kneib2018Scission,Longari2023CANdito}, with a focus on specific domains.

However, many previous solutions either addressed the problem theoretically or via a pure software approach, without considering the peculiarities of the automotive context in their design. While the performance evaluation of security solutions for in-vehicle networks can be easily demonstrated via a pure software perspective, proving the vulnerabilities affecting existing automotive components or protocols requires access to a real vehicle system, involving investments that might prevent many researchers from undertaking this crucial research activity.

In this work, we present \textit{HackCar}, a test platform for attacks and defenses on a cost-contained automotive architecture. Our platform is designed to be fully extensible and configurable for every scenario that security researchers want to address, from in-vehicle network security to V2X communications. We based the design of \textit{HackCar} on an $F1-10^{th}$ model stripped of all unnecessary components, on which we built our computing platform. To prove the effectiveness of our design, we compared the utilization of the in-vehicle network deployed on the \textit{HackCar} test platform and we implemented a forward-collision avoidance system that, coupled with a simple autonomous driving algorithm, allows the prototype to stop in case an obstacle is detected by the forward-facing sensors. Then, we replicated the effects of an attack to the autonomous driving system, resulting in the platform not stopping in case an obstacle is detected, thus crashing with it.

The main contribution to the state-of-the-art of \textit{HackCar} is the public release of the specifications, designs, and prototype boards required to fully implement our test system. This allows future researchers to replicate our prototype, expand on its capabilities, and build a more comprehensive test platform for attacks and defenses on a secure, safe, and budget-friendly automotive system.

\subsection{Related Work}
\label{ss:related}

\textbf{Attacks on the automotive systems.} The specific challenges posed by the automotive context in attacking vehicle systems have been extensively discussed in the literature~\cite{Checkoway2011Comprehensive}. One of the most influential works in this field is presented in~\cite{MillerValasekBlackHatPaper}, where the authors demonstrated a practical approach to injecting malicious messages into CAN communication, enabling the remote takeover of a vehicle by bypassing the driver's commands. Since then, numerous efforts have been made by security practitioners to introduce novel attacks on automotive systems~\cite{Lee2015Fuzzing, Cho2016Error, Palanca2017Stealth,KeenSecurity}, although these efforts often require access to a physical vehicle for result replication.

\textbf{Defenses for the automotive systems.} Immediately following the initial public demonstrations of attacks on automotive systems, numerous security researchers began addressing the issue by designing novel detection algorithms for in-vehicle communication. These algorithms exploit either the physical characteristics of the microcontrollers~\cite{Cho2017Viden,Kneib2018Scission}, the specifications of the CAN bus protocol~\cite{Gmiden2016Detector, Stabili2019Missing}, or the aggregation of metadata at the network level~\cite{nowdehi2019casad}. Many of these solutions rely on batch analysis of the content of CAN communication to scrutinize the data. While this approach is adequate for showcasing the detection performance of the solution, it falls short of meeting the specifications of automotive microcontrollers. However, this trend appears to be on the decline, as many recent works published in the last couple of years discuss the application of their solutions on different microcontrollers~\cite{Stabili2022DAGA,Longari2023CANdito}.

\textbf{Cyber Test platform for automotive.} Benchmarking poses one of the significant challenges for the automotive industry~\cite{kramer2015real}. Many solutions are customized by individual car manufacturers for their proprietary systems, and portability is severely limited. From benchmarks for complex systems requiring a dedicated simulation environment~\cite{elmqvist2003real, buranathiti2005benchmark, gangel2021modelling, Paranjape2020Simulation, Wallace2022Simulation, Qiao2023Simulation} to benchmarks designed for hardware-in-the-loop scenarios~\cite{ellies2016benchmarking,shao2019evaluating, abboush2022hardware, Pechinger2020HITL, Scheffe2023Hybrid}, various approaches can be adopted. However, each approach necessitates a dedicated setup and has limited applicability.
In the realm of benchmarks for anomaly detectors in in-vehicle communication, researchers have started presenting generic frameworks for comparing the detection performance of algorithms~\cite{Stabili2021ITASEC}. They also provide a targeted analysis of detectors that share the same detection metric~\cite{Pollicino2023Performance} and address this issue from a multidisciplinary perspective~\cite{Stabili2023Multidisciplinary}. However, these solutions rely on a simulated environment for testing detection algorithms, without considering the specific context of the automotive industry or providing a platform for replicating the consequences of cyber attacks on automotive systems.
In this work, we aim to address all these issues by presenting a physical prototype for benchmarking both attacks and defenses on a real automotive system.

\subsection{Outline}
\label{ss:outline}
The remainder of the paper is organized as follows. Section~\ref{s:design} outlines the requirements and the design of \textit{HackCar}, while Section~\ref{s:implementation} presents the implementation and the components selected for our prototype. Section~\ref{experimental_evaluation} discusses the validity of the implemented prototype in comparison to a real vehicle system and demonstrates the consequences of the attack on the test vehicle. Finally, conclusions and future research directions are addressed in Section~\ref{s:conclusion}.

%% file: sections/implementation.tex
\section{Design of the platform}
\label{s:design}

In this section, we present the requirements (\ref{ss:requirements}) and the design (\ref{ss:platform_design}) of the \emph{HackCar} test platform. Specifically, we will discuss the design of the three main components of the test platform: (I) the sensing system, (II) the on-board controller, and (III) the in-vehicle network. Although each of these components is discussed as a single entity in our design, during the implementation process, we emphasize that a single component can be composed of multiple elements without impacting the previously made design choices.

\subsection{Requirements of the platform}
\label{ss:requirements}
To define the requirements of the test platform, it is necessary to first outline the operational scenarios that we aim to support. Given that the primary objective of this work is to present a platform for analyzing the consequences of cyber-attacks on the system under various circumstances, the initial scenario we define supports self-driving capabilities. It has the ability to move along a pre-defined route without requiring an external controller, featuring an enabled Automatic Emergency Braking (AEB) system, thereby replicating the functionalities of a \emph{level 3+} Advanced Driver Assistance System (ADAS).
Building upon this operational scenario, the second one is defined by disabling self-driving capabilities and introducing an external controller to command the platform. In this second scenario, the same AEB system is deployed as in the first scenario, providing basic \emph{level 1} ADAS capabilities.

In summary, the two operational scenarios supported by the \textit{HackCar} platform are:
\begin{enumerate}
    \item \textbf{ManualAEB}, where the platform is manually controlled by a remote controller, and an Automatic Emergency Braking (AEB) system is enabled to replicate a \emph{level 1} ADAS.
    \item \textbf{AutoDrive}, where the platform can move along a pre-defined route without requiring any external controller and with the AEB system enabled, replicating the functionality of a \emph{level 3+} ADAS.
\end{enumerate}

To support these two operational scenarios, the requirements of the test platform (assuming that the actuators to replicate the behavior of a vehicle by accelerating, steering, and braking are supported) can be summarized as follows:

\begin{itemize}
    \item Wireless interface for the remote controller;
    \item Sensing system for forward-object detection;
    \item On-board controller to operate on the braking system (for both scenarios) and the accelerating system (in the \emph{AutoDrive} scenario);
    \item On-board controller with the ability to switch between the two operational modes.
\end{itemize}

Additionally, we include two dedicated microcontrollers to support different attacks and defense solutions at n-vehicle network level. Despite these microcontrollers enable researchers to deploy attacks and defense solutions to the \emph{HackCar} platform, we emphasize that this paper does not focus on discussing novel attack scenarios or novel detection techniques. The inclusion of support for both attacks and defense mechanisms in the design process of the test platform is representative of its capabilities for demonstrating attack consequences and testing security solutions in the automotive context without requiring a full-size vehicle.

\subsection{Design of the platform}
\label{ss:platform_design}
\begin{figure}[hptb]
    \centering
    \includegraphics[width=.9\columnwidth]{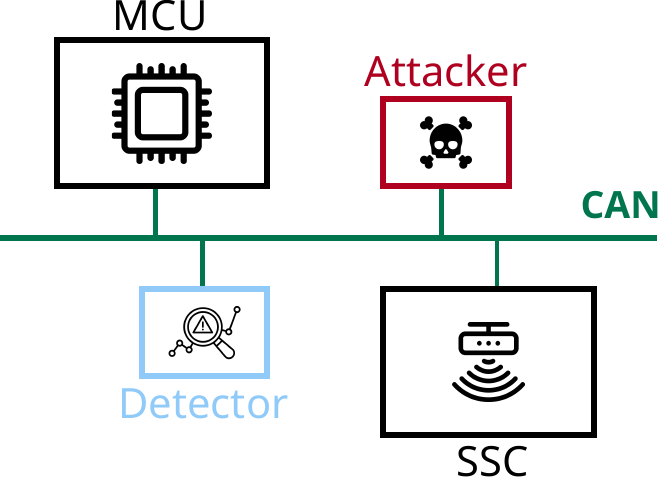}
    \caption{High-level overview of the \emph{HackCar} test platform}
    \label{fig:hackcar_overview_design}
\end{figure}

The \emph{HackCar} test platform is constructed with three main components (the \emph{sensing system}, the \emph{on-board controllers} and the \emph{in-vehicle network}), each responsible for distinct functions to replicate the operational scenarios supported by the platform. A high-level representation of the design of the \emph{HackCar} test platform is shown in Figure~\ref{fig:hackcar_overview_design}.

\emph{Sensing system design.} To facilitate the development of operational scenarios, we require a sensing system that supports at least \emph{level 3} self-driving capabilities, with a primary emphasis on the obstacle detection system. Despite defining both \emph{ManualAEB} and \emph{AutoDrive} operational scenarios as representative of two different ADAS levels, it is noteworthy that the only sensing system required for both scenarios is the obstacle detection system. The forward obstacle detection feature of \emph{HackCar} can be implemented using either a LiDAR-based solution or stereo cameras directed towards the front of the test platform.

\emph{On-board controller.} The on-board controller serves as the computational core of our system, analyzing data generated by the sensing system to detect obstacles, halt the prototype, and prevent collisions. Consequently, the on-board controller is tasked with managing the actuators that control the platform's dynamics, including the braking system (for both operational scenarios) and the accelerating and steering systems (only for the \emph{AutoDrive} scenario). Additionally, the on-board controller must provide a wireless interface for the remote controller, allowing input in the \emph{ManualAEB} scenario.

Given the diverse features required at the controller level, it is advisable to design the test platform with a multi-controller approach instead of opting for a single-board controller. In our final design, we have incorporated four different controllers:
\begin{itemize}
    \item \textbf{Sensing system controller (SSC):} Responsible for evaluating the presence of an obstacle in the proximity of the vehicle by analyzing data generated from the sensing system.
    \item \textbf{Main controller unit (MCU):} Responsible for accessing the actuators of the test platform and managing the two operational scenarios. Specifically, when a specific operational scenario is selected, the main controller is responsible for implementing all the required functions. In our prototype, the main controller provides the wireless interface used to connect the remote controller to the platform (in the \emph{ManualAEB} scenario). It is also responsible for implementing the self-driving capabilities required by the \emph{AutoDrive} scenario.
    \item \textbf{Attack controller (Attacker):} Responsible for replicating attacks on the vehicle to study the consequences of each attack scenario in different operational scenarios.
    \item \textbf{Detection controller (Detector):} Responsible for detecting anomalies in the prototype's behavior by analyzing various types of data according to the tested detection algorithm.
\end{itemize}

\emph{In-Vehicle Networks.} To facilitate communication among the controllers that constitute the computational platform of our test prototype, we designed them to communicate with each other through an in-vehicle network, replicating the behavior of a full-sized vehicle. Therefore, the obvious choice in designing the in-vehicle network communication was to rely on the Controller Area Network (CAN), as it is the most widely deployed communication network for in-vehicle applications. The CAN is an industrial bus standard designed to facilitate data exchange between devices without requiring a host computer. It is widely adopted in the automotive context due to its high resilience to electromagnetic interference and cost-effectiveness.
Data on the CAN bus is exchanged by microcontrollers via \emph{data frames}, a specific type of message defined in the CAN standard. The two most crucial fields of the data frame are the \emph{ID} (used to identify the type of message) and the \emph{DATA} (a sequence of up to $64$ bits containing various values). Each data frame with a particular ID value is sent by one and only one microcontroller, while its content can be used by many microcontrollers for their functioning. As the CAN standard does not define how signals are encoded in the data field, it is necessary to define the number, structure, encoding, and semantics of the encoded signals.

\section{Implementation of the platform}
\label{s:implementation}
In this section, we discuss the architectural decisions made during the implementation of the three core components of our platform. The whole implementation process is based on a Traxxas Ford Fiesta ST Rally~\cite{traxxas} model as the foundational platform for developing the sensing, controller, and communication systems. The chosen platform is a scaled-down replica, maintaining a $1:10$ ratio compared to the original Ford Fiesta ST Rally vehicle. This model is equipped with all the necessary actuators to access braking, accelerating, and steering functions.

\emph{Sensing system.} The sensing system is built around a Hokuyo $UST-10LX$ $2D$ LiDAR~\cite{Houko2DLiDAR}. This LiDAR is widely utilized for implementing detection and localization in autonomous mobile robots, automated guided vehicles, and carts. Equipped with an Ethernet interface, it offers a $270\degree$ field-of-view up to a distance of $10$ meters. With $1081$ measurement steps, it achieves an angular resolution of $0.25\degree$ and boasts a rapid response time of $25ms$, making it one of the most reliable components for object detection.

\emph{On-Board Controllers.} The on-board controllers are implemented on different microcontrollers to replicate the heterogeneity of a typical in-vehicle electrical and electronic architecture. The details are presented as follows:
\begin{itemize}
    \item \textbf{Sensing system controller (SSC).} The SSC is implemented on a NVIDIA Jetson Nano board~\cite{JetsonNano}, a single-board computer designed for low-power environments and equipped with Robot Operating System (ROS) 2~\cite{ROS2}. Initially developed for robotic systems, ROS2's high versatility led to its application in real-time systems as well. It is important to note that the SSC is configured to provide only basic self-driving and object detection capabilities, although the current hardware version of \emph{HackCar} can support more advanced and full self-driving capabilities.
    \item \textbf{Main controller unit (MCU).} The MCU is implemented on an Infineon Aurix TriCore $TC297$~\cite{InfineonAurix} microcontroller, an integrated and advanced automotive-grade microcontroller widely adopted in the automotive industry. The Aurix TriCore controller is connected to the actuators of the Traxxas model to control braking, acceleration, and steering via a Trampa VESC $6 MKVI$ Electronic Speed Controller (ESC)~\cite{VESC}. The VESC ESC allows us to control all the systems responsible for managing the platform's dynamics via a single interface, thereby reducing the complexity of the different sub-systems.
    \item \textbf{Attack controller.} The attack controller is implemented on a CANPico board~\cite{CANPico}, equipped with an $MPC2517FD$ CAN controller and a CAN transceiver to provide access to the CAN bus. The CANPico is shipped with a source SDK for Micropython, supporting the CANHack toolkit API~\cite{CANHack} for low-level attacks on the CAN bus.
    \item \textbf{Detection controller.} The detection controller is based on another Infineon Aurix TriCore $TC297$ controller. We emphasize that the performance evaluation of various anomaly detectors for CAN networks is beyond the scope of this work. However, we chose to implement a detection controller on the Aurix TriCore board due to the interfaces available on the board, making it a valid choice for designing and testing solutions that go beyond the only CAN bus.
\end{itemize}

\emph{In-vehicle CAN.} In-vehicle communication relies on CAN messages transmitted every $10ms$ from various controllers. Since the CAN protocol allows car manufacturers to define the content and encoding of signals on the CAN bus, we implemented our network based on these messages
\begin{itemize}
    \item \texttt{RPM [ID: 0x400, DLC: 4 Bytes]}: This message contains the target RPMs of the engine in the \textit{AutoDrive} operational scenario. The \textit{sensing system controller} generates this message, while the \textit{main controller} uses it to set the RPMs of the engine via the VESC controller.
    \item \texttt{STEERING [ID: 0x401, DLC: 4 Bytes]}: This message holds the target radial value of the steering wheel required in both the \textit{ManualAEB} (converted from user input) and \textit{AutoDrive} scenarios. The \textit{sensing system controller} generates this message, and the \textit{main controller} uses it to set the angle of the steering wheel via the VESC controller.
    \item \texttt{BREAK [ID: 0x402, DLC: 1 Byte]}: This message contains the target value of the braking system required for the \textit{Automatic Emergency Brake} feature. The \textit{sensing system controller} generates this message, and the \textit{main controller} uses it to decrease the value of the RPM of the engine via the VESC controller.
\end{itemize}
In addition to these messages, we implemented three \texttt{service} messages to switch between the two operational scenarios (CAN ID \texttt{0x500}), activate and deactivate the AEB (CAN ID \texttt{0x501}), and trigger the execution of attacks on the \textit{Attack controller} (CAN ID \texttt{0x502}). These messages are defined with a DLC of $1$ Byte and are only sent on the network when required by the user.

%% file: sections/experimental_evaluation.tex
\section{Experimental evaluation}
\label{experimental_evaluation}

In this section, we present the experimental evaluation of the HackCar test platform. Before presenting the results of the evaluation, we introduce the attacks currently supported by the platform in Section~\ref{ss:threat_model}, while both validation and functional evaluation are discussed in Section~\ref{ss:val_funct}. 

\subsection{Threat model}
\label{ss:threat_model}
The threat model supported in the prototype version of HackCar is based on the most effective type of attacks for in-vehicle communication, namely the \emph{message injection} attack. Specifically, we consider an attacker with access to the in-vehicle network through either the On-Board Diagnostic interfaces, an additional device maliciously installed on the vehicle, or by compromising an ECU~\cite{MillerValasekBlackHatPaper}. We do not consider the scenario where the attacker is able to compromise an existing ECU to modify its normal behavior.

In our experimental evaluation, we assume the attacker can subvert the normal behavior of the AEB system deployed on the \emph{sensing system controller}. For this purpose, the attacker controller is configured to intercept the \texttt{RPM} message on the CAN, preventing the main controller from stopping the platform. To replicate a stealthy behavior, the attacker controller is configured to send the message \emph{after} the one generated by the \emph{sensing system controller} is sent on the network, effectively overwriting it.

\subsection{Validation and Functional evaluation}
\label{ss:val_funct}
For the experimental evaluation of the \emph{HackCar} test platform, we focus on the experimental validation of the platform concerning the in-vehicle network of an unmodified, licensed vehicle and the functional evaluation of the platform against the attack discussed in the threat model. We emphasize that, despite \emph{HackCar} offering support for a defense system, this work does not focus on the detection capabilities since it would only require the implementation of existing IDS for CAN communication~\cite{IV2017,Longari2023CANdito,Pollicino2023Performance}, thus not providing any substantial contribution to the state-of-the-art in this particular field.

\textbf{HackCar Validation.} The validation of the HackCar test platform is based on the analysis of CAN bus utilization concerning the in-vehicle data collected from an unmodified, licensed vehicle. The objective of this test is to compare our implementation with the most frequently messages in our reference CAN data, as these are likely to be messages carrying sensor readings related to the \emph{drive-by-wire} features of our reference vehicle.

\textbf{Functional Evaluation.} In the functional validation of the HackCar test platform, we focus on analyzing the content of CAN messages under attack to scrutinize the behavior of the test platform while replicating the attack described in the previous section. The aim of this test is to demonstrate that, under attack conditions, the consequences of the attack can also be observed at the CAN communication level.
 
\subsection{Results analysis}
\label{ss:results}

\begin{figure}[hpbt]
    \centering
    \includegraphics[width=.9\columnwidth]{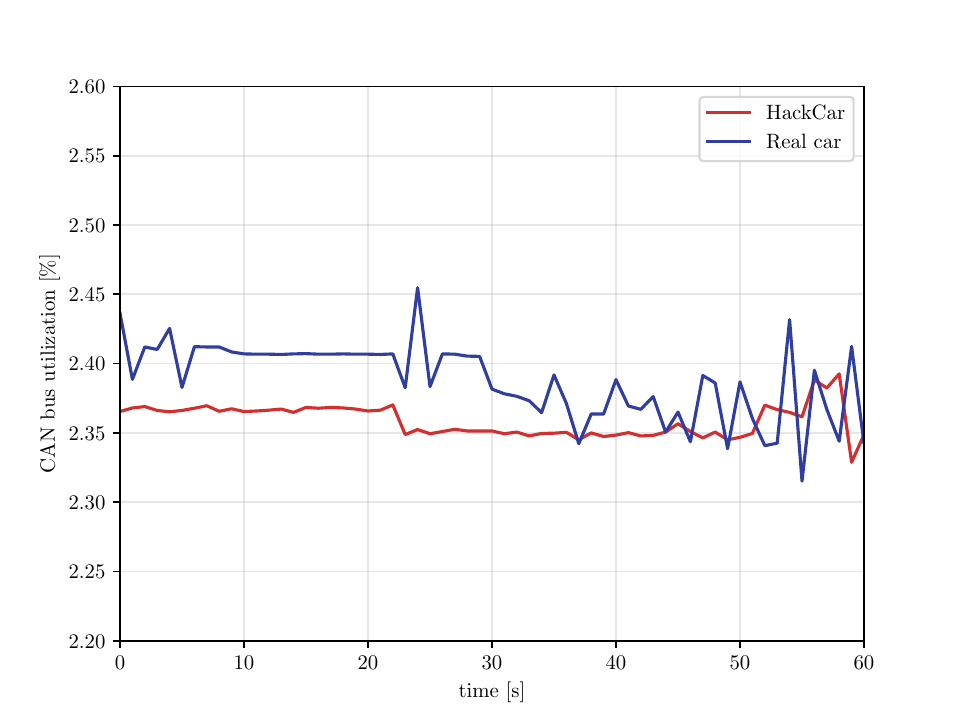}
    \caption{Validation of the HackCar test platform against a real vehicle}
    \label{fig:utilization_validation}
\end{figure}

\textbf{HackCar Validation Results.} The results of the validation tests are presented in Figure~\ref{fig:utilization_validation}, where the CAN bus utilization of the test platform is compared with data collected from an unmodified, licensed vehicle.

Figure~\ref{fig:utilization_validation} compares the CAN bus utilization (represented on the $y$-axis) of the HackCar test platform (red line) with the most frequent messages (sent with a frequency of $100$ Hz) of our reference vehicle (blue line) over a duration of $60$ seconds of CAN data. The results presented in Figure~\ref{fig:utilization_validation} indicate that both platforms exhibit similar behavior, with the HackCar test platform reaching approximately the same utilization percentage as the reference vehicle. We emphasize that these results serve as a simple yet effective demonstration that the implemented architecture can be configured to behave similarly to a specific subset of real vehicle CAN messages responsible for the same functions.

\begin{figure}[hpbt]
    \centering
    \includegraphics[width=.9\columnwidth]{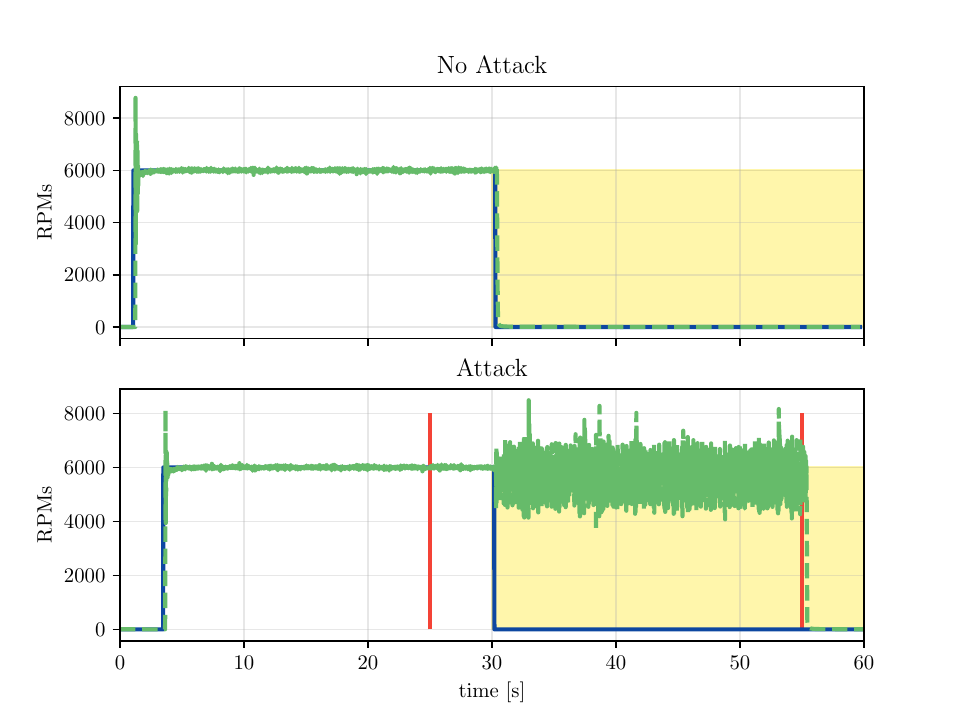}
    \caption{Functional evaluation of the HackCar test platform in a normal operating scenario (top plot) and under attack (bottom plot)}
    \label{fig:functional_evaluation}
\end{figure}

\textbf{Functional evaluation results.} The results of the functional evaluation test are presented in Figure~\ref{fig:functional_evaluation}, where the rotational speed of the engine is compared with the target speed set by the controller.

Figure~\ref{fig:functional_evaluation} illustrates the behavior of the HackCar test platform in both an attack-free (top plot) and under-attack  (bottom plot) scenarios. In both scenarios, the target \textit{rotation per minute} (RPMs) set by the MCU are depicted in blue, while the actual RPMs of the HackCar platform, as observed from the electric engine, are represented by the green dashed line. The yellow cross-hatched area in both plots indicates the presence of an obstacle in front of the HackCar, while the red vertical lines in the bottom plot denote the start and end of the attack.

In both plots, the obstacle is detected approximately $0.5$ meters in front of the vehicle after around $30$ seconds, with the attack triggered $5$ seconds before the detection of the obstacle in the second scenario. Analyzing the results presented in Figure~\ref{fig:functional_evaluation}, it is evident that there is an initial oscillation in the real RPMs due to the action of the HackCar platform controller, resulting in reaching the target speed of $6000$ RPMs a few seconds after the start of the HackCar. The speed is then maintained around the equilibrium point ($6000$ RPMs) until the obstacle is detected.

In the scenario depicted in the top part of Figure~\ref{fig:functional_evaluation}, the detection of the obstacle in front of the HackCar results in the MCU setting the target speed to $0$ RPMs, forcing the vehicle to stop in under a second. On the other hand, in the second scenario, the actual RPMs of the electric engine are maintained around the equilibrium point by the controller because the attack is triggered before the detection of the obstacle. In this latter scenario, it is also noticeable that the actual RPMs are not as stable as before the obstacle detection, owing to the attack targeting the controller. The attack introduces sensible oscillations in the system since, apart from receiving malicious messages containing the target speed of $6000$ RPMs, the controller also receives legitimate messages from the MCU containing the target speed of $0$ RPMs. This behavior, as already showcased in~\cite{Stabili2022Consequences}, is a realistic representation of the normal operation of an engine controller.

%% file: sections/conclusions.tex
\section{Conclusion}
\label{s:conclusion}

In this paper, we presented the design of \textit{HackCar}, a test platform for attacks and defenses on a generic and configurable automotive system. We supported our design with a real implementation based on an \emph{F1-$10^{th}$} platform, equipped with at least $4$ different automotive-grade microcontrollers. These microcontrollers are responsible for managing the sensing system of the platform, enabling the implementation of various ADAS features. The main controller governs the target speed and heading of the vehicle while operating in autonomous mode. The platform also includes a malicious attacker with access to in-vehicle communication, providing researchers the ability to prototype novel attacks on a small scale or replicate known cyberattacks to study their consequences in a realistic automotive environment. Additionally, a detection system is incorporated, enabling researchers to develop and test architectural solutions to existing or novel cyberattacks on the vehicle system.

The results of the validation and functional behavior tests demonstrate that our platform behaves similarly to an unmodified licensed vehicle. The system can be easily utilized by security researchers to showcase the consequences of cyberattacks on an automotive system without requiring access to a full vehicle.

%% file: main.bbl
\begin{thebibliography}{10}
\providecommand{\url}[1]{#1}
\csname url@samestyle\endcsname
\providecommand{\newblock}{\relax}
\providecommand{\bibinfo}[2]{#2}
\providecommand{\BIBentrySTDinterwordspacing}{\spaceskip=0pt\relax}
\providecommand{\BIBentryALTinterwordstretchfactor}{4}
\providecommand{\BIBentryALTinterwordspacing}{\spaceskip=\fontdimen2\font plus
\BIBentryALTinterwordstretchfactor\fontdimen3\font minus \fontdimen4\font\relax}
\providecommand{\BIBforeignlanguage}[2]{{%
\expandafter\ifx\csname l@#1\endcsname\relax
\typeout{** WARNING: IEEEtran.bst: No hyphenation pattern has been}%
\typeout{** loaded for the language `#1'. Using the pattern for}%
\typeout{** the default language instead.}%
\else
\language=\csname l@#1\endcsname
\fi
#2}}
\providecommand{\BIBdecl}{\relax}
\BIBdecl

\bibitem{BOSCHCanSpecV2}
Bosch. (1991) Can specification version 2.0.

\bibitem{MillerValasekAdventures}
C.~Miller and C.~Valasek. (2014) Adventures in automotive networks and control units.

\bibitem{KeenSecurity}
\BIBentryALTinterwordspacing
{Keen Security Lab of Tencent}. (2016) Car hacking research: Remote attack tesla motors. [Online]. Available: \url{http://keenlab.tencent.com/en/2016/09/19/Keen-Security-Lab-of-Tencent-Car-Hacking-Research-Remote-Attack-to-Tesla-Cars/}
\BIBentrySTDinterwordspacing

\bibitem{Kneib2018Scission}
\BIBentryALTinterwordspacing
M.~Kneib and C.~Huth, ``Scission: Signal characteristic-based sender identification and intrusion detection in automotive networks,'' in \emph{Proceedings of the 2018 ACM SIGSAC Conference on Computer and Communications Security}, ser. CCS '18.\hskip 1em plus 0.5em minus 0.4em\relax New York, NY, USA: ACM, 2018, pp. 787--800. [Online]. Available: \url{http://doi.acm.org/10.1145/3243734.3243751}
\BIBentrySTDinterwordspacing

\bibitem{Stabili2022DAGA}
D.~Stabili, L.~Ferretti, M.~Andreolini, and M.~Marchetti, ``Daga: Detecting attacks to in-vehicle networks via n-gram analysis,'' \emph{IEEE Transactions on Vehicular Technology}, vol.~10, p.~15, 2022.

\bibitem{Groza2012LibraCAN}
B.~Groza, S.~Murvay, A.~van Herrewege, and I.~Verbauwhede, ``Libra-can: A lightweight broadcast authentication protocol for controller area networks,'' in \emph{Cryptology and Network Security}, J.~Pieprzyk, A.-R. Sadeghi, and M.~Manulis, Eds.\hskip 1em plus 0.5em minus 0.4em\relax Berlin, Heidelberg: Springer Berlin Heidelberg, 2012, pp. 185--200.

\bibitem{Kurachi2016CaCAN}
R.~Kurachi, Y.~Matsubara, H.~Takada, H.~Ueda, and S.~Horihata, ``Cacan: Centralized authentication system in can (controller area network),'' 2016.

\bibitem{MillerValasekBlackHatPaper}
\BIBentryALTinterwordspacing
C.~Miller and C.~Valasek. (2015) Remote exploitation of an unaltered passenger vehicle. White paper of Blackhat US conference. [Online]. Available: \url{http://illmatics.com/Remote\%20Car\%20Hacking.pdf}
\BIBentrySTDinterwordspacing

\bibitem{Longari2023CANdito}
S.~Longari, C.~A. Pozzoli, A.~Nichelini, M.~Carminati, and S.~Zanero, ``Candito: Improving payload-based detection of attacks on controller area networks,'' in \emph{Cyber Security, Cryptology, and Machine Learning}, S.~Dolev, E.~Gudes, and P.~Paillier, Eds.\hskip 1em plus 0.5em minus 0.4em\relax Cham: Springer Nature Switzerland, 2023, pp. 135--150.

\bibitem{Checkoway2011Comprehensive}
\BIBentryALTinterwordspacing
S.~Checkoway, D.~McCoy, B.~Kantor, D.~Anderson, H.~Shacham, S.~Savage, K.~Koscher, A.~Czeskis, F.~Roesner, and T.~Kohno, ``Comprehensive experimental analyses of automotive attack surfaces,'' in \emph{Proceedings of the 20th USENIX Conference on Security}, ser. SEC'11.\hskip 1em plus 0.5em minus 0.4em\relax Berkeley, CA, USA: USENIX Association, 2011, pp. 6--6. [Online]. Available: \url{http://dl.acm.org/citation.cfm?id=2028067.2028073}
\BIBentrySTDinterwordspacing

\bibitem{Lee2015Fuzzing}
H.~{Lee}, K.~{Choi}, K.~{Chung}, J.~{Kim}, and K.~{Yim}, ``Fuzzing {CAN} packets into automobiles,'' in \emph{2015 IEEE 29th Int'l Conf. on Advanced Information Networking and Applications}, March 2015.

\bibitem{Cho2016Error}
\BIBentryALTinterwordspacing
K.-T. Cho and K.~G. Shin, ``Error handling of in-vehicle networks makes them vulnerable,'' in \emph{Proceedings of the 2016 ACM SIGSAC Conference on Computer and Communications Security}, ser. CCS '16.\hskip 1em plus 0.5em minus 0.4em\relax New York, NY, USA: ACM, 2016, pp. 1044--1055. [Online]. Available: \url{http://doi.acm.org/10.1145/2976749.2978302}
\BIBentrySTDinterwordspacing

\bibitem{Palanca2017Stealth}
\BIBentryALTinterwordspacing
A.~Palanca, E.~Evenchick, F.~Maggi, and S.~Zanero, ``A stealth, selective, link-layer denial-of-service attack against automotive networks,'' in \emph{Detection of Intrusions and Malware, and Vulnerability Assessment - 14th International Conference, {DIMVA} 2017, Bonn, Germany, July 6-7, 2017, Proceedings}, 2017, pp. 185--206. [Online]. Available: \url{https://doi.org/10.1007/978-3-319-60876-1\_9}
\BIBentrySTDinterwordspacing

\bibitem{Cho2017Viden}
\BIBentryALTinterwordspacing
K.~Cho and K.~G. Shin, ``Viden: Attacker identification on in-vehicle networks,'' \emph{CoRR}, vol. abs/1708.08414, 2017. [Online]. Available: \url{http://arxiv.org/abs/1708.08414}
\BIBentrySTDinterwordspacing

\bibitem{Gmiden2016Detector}
M.~{Gmiden}, M.~H. {Gmiden}, and H.~{Trabelsi}, ``{An intrusion detection method for securing in-vehicle CAN bus},'' in \emph{Int'l Conf. Sciences and Techniques of Automatic Control and Computer Engineering}, 2016.

\bibitem{Stabili2019Missing}
D.~{Stabili} and M.~{Marchetti}, ``Detection of missing {CAN} messages through inter-arrival time analysis,'' in \emph{2019 IEEE 90th Vehicular Technology Conf.}, Sep. 2019.

\bibitem{nowdehi2019casad}
N.~{Nowdehi}, W.~{Aoudi}, M.~{Almgren}, and T.~{Olovsson}, ``{CASAD}: {CAN}-aware stealthy-attack detection for in-vehicle networks,'' 2019.

\bibitem{kramer2015real}
S.~Kramer, D.~Ziegenbein, and A.~Hamann, ``Real world automotive benchmarks for free,'' in \emph{6th International Workshop on Analysis Tools and Methodologies for Embedded and Real-time Systems (WATERS)}, vol. 130, 2015.

\bibitem{elmqvist2003real}
H.~Elmqvist, S.~E. Mattsson, H.~Olsson, J.~Andreasson, M.~Otter, C.~Schweiger, and D.~Br{\"u}ck, ``Real-time simulation of detailed automotive models,'' in \emph{Proceedings}, 2003, pp. 29--38.

\bibitem{buranathiti2005benchmark}
T.~Buranathiti and J.~Cao, ``Benchmark simulation results: automotive underbody cross member (benchmark 2),'' in \emph{AIP Conference Proceedings}, vol. 778, no.~1.\hskip 1em plus 0.5em minus 0.4em\relax American Institute of Physics, 2005, pp. 1004--1112.

\bibitem{gangel2021modelling}
K.~Gangel, Z.~Hamar, A.~H{\'a}ry, {\'A}.~Horv{\'a}th, G.~Jand{\'o}, B.~K{\"o}nyves, D.~Panker, K.~Pint{\'e}r, M.~Pataki, M.~Szalai \emph{et~al.}, ``Modelling the zalazone proving ground: a benchmark of state-of-the-art automotive simulators prescan, ipg carmaker, and vtd vires,'' \emph{Acta Technica Jaurinensis}, vol.~14, no.~4, pp. 488--507, 2021.

\bibitem{Paranjape2020Simulation}
I.~Paranjape, A.~Jawad, Y.~Xu, A.~Song, and J.~Whitehead, ``A modular architecture for procedural generation of towns, intersections and scenarios for testing autonomous vehicles,'' in \emph{2020 IEEE Intelligent Vehicles Symposium (IV)}, 2020, pp. 162--168.

\bibitem{Wallace2022Simulation}
A.~Wallace, S.~Khastgir, X.~Zhang, S.~Brewerton, B.~Anctil, P.~Burns, D.~Charlebois, and P.~Jennings, ``Validating simulation environments for automated driving systems using 3d object comparison metric,'' in \emph{2022 IEEE Intelligent Vehicles Symposium (IV)}, 2022, pp. 860--866.

\bibitem{Qiao2023Simulation}
Z.~Qiao, X.~Sun, H.~Loeb, and R.~Mangharam, ``Drive right: Shaping public’s trust, understanding, and preference towards autonomous vehicles using a virtual reality driving simulator,'' in \emph{2023 IEEE Intelligent Vehicles Symposium (IV)}, 2023, pp. 1--8.

\bibitem{ellies2016benchmarking}
B.~Ellies, C.~Schenk, and P.~Dekraker, ``Benchmarking and hardware-in-the-loop operation of a 2014 mazda skyactiv 2.0 l 13: 1 compression ratio engine,'' SAE Technical Paper, Tech. Rep., 2016.

\bibitem{shao2019evaluating}
Y.~Shao, M.~A.~M. Zulkefli, Z.~Sun, and P.~Huang, ``Evaluating connected and autonomous vehicles using a hardware-in-the-loop testbed and a living lab,'' \emph{Transportation Research Part C: Emerging Technologies}, vol. 102, pp. 121--135, 2019.

\bibitem{abboush2022hardware}
M.~Abboush, D.~Bamal, C.~Knieke, and A.~Rausch, ``Hardware-in-the-loop-based real-time fault injection framework for dynamic behavior analysis of automotive software systems,'' \emph{Sensors}, vol.~22, no.~4, p. 1360, 2022.

\bibitem{Pechinger2020HITL}
M.~Pechinger, G.~Schröer, K.~Bogenberger, and C.~Markgraf, ``Hardware in the loop test using infrastructure based emergency trajectories for connected automated driving,'' in \emph{2020 IEEE Intelligent Vehicles Symposium (IV)}, 2020, pp. 357--362.

\bibitem{Scheffe2023Hybrid}
P.~Scheffe and B.~Alrifaee, ``A scaled experiment platform to study interactions between humans and cavs,'' in \emph{2023 IEEE Intelligent Vehicles Symposium (IV)}, 2023, pp. 1--6.

\bibitem{Stabili2021ITASEC}
D.~Stabili, F.~Pollicino, and A.~Rota, ``A benchmark framework for can ids,'' in \emph{Italian Conference on Cybersecurity 2021 (ITASEC)}, April 2021, pp. 233--245.

\bibitem{Pollicino2023Performance}
\BIBentryALTinterwordspacing
F.~Pollicino, D.~Stabili, and M.~Marchetti, ``Performance comparison of timing-based anomaly detectors for controller area network: A reproducible study,'' \emph{ACM Trans. Cyber-Phys. Syst.}, jun 2023, just Accepted. [Online]. Available: \url{https://doi.org/10.1145/3604913}
\BIBentrySTDinterwordspacing

\bibitem{Stabili2023Multidisciplinary}
\BIBentryALTinterwordspacing
D.~Stabili, R.~Romagnoli, M.~Marchetti, B.~Sinopoli, and M.~Colajanni, ``A multidisciplinary detection system for cyber attacks on powertrain cyber physical systems,'' \emph{Future Generation Computer Systems}, vol. 144, pp. 151--164, 2023. [Online]. Available: \url{https://www.sciencedirect.com/science/article/pii/S0167739X23000602}
\BIBentrySTDinterwordspacing

\bibitem{traxxas}
\BIBentryALTinterwordspacing
{Traxxas}. (2024) {Ford Fiesta ST Rally 1:10 model}. [Online]. Available: \url{https://traxxas.com/products/models/electric/ford-fiesta-st-rally}
\BIBentrySTDinterwordspacing

\bibitem{Houko2DLiDAR}
\BIBentryALTinterwordspacing
{Hokuyo Automatic USA Corporation}. (2024) {UST-10LX}. [Online]. Available: \url{https://hokuyo-usa.com/products/lidar-obstacle-detection/ust-10lx}
\BIBentrySTDinterwordspacing

\bibitem{JetsonNano}
\BIBentryALTinterwordspacing
{NVIDIA}. (2024) {Jetson Nano Developer Kit}. [Online]. Available: \url{https://developer.nvidia.com/embedded/jetson-nano-developer-kit}
\BIBentrySTDinterwordspacing

\bibitem{ROS2}
\BIBentryALTinterwordspacing
{Open Robotics}. (2024) {Robot Operating System v2}. [Online]. Available: \url{https://github.com/ros2}
\BIBentrySTDinterwordspacing

\bibitem{InfineonAurix}
\BIBentryALTinterwordspacing
{Infineon}. (2024) {AURIX Family - TC297TA}. [Online]. Available: \url{https://www.infineon.com/cms/en/product/microcontroller/32-bit-tricore-microcontroller/32-bit-tricore-aurix-tc2xx/aurix-family-tc297ta-adas/}
\BIBentrySTDinterwordspacing

\bibitem{VESC}
\BIBentryALTinterwordspacing
{TRAMPA Boards}. (2024) {VESC 6 MkVI}. [Online]. Available: \url{https://trampaboards.com/vesc-6-mkvi--the-amazing-trampa-vesc-6-mkvi--gives-maximum-power-original-p-27536.html}
\BIBentrySTDinterwordspacing

\bibitem{CANPico}
\BIBentryALTinterwordspacing
{CANIS Automotive Labs}. (2024) {CANPico}. [Online]. Available: \url{https://canislabs.com/canpico/}
\BIBentrySTDinterwordspacing

\bibitem{CANHack}
\BIBentryALTinterwordspacing
------. (2024) {CANHack}. [Online]. Available: \url{https://github.com/kentindell/canhack}
\BIBentrySTDinterwordspacing

\bibitem{IV2017}
M.~{Marchetti} and D.~{Stabili}, ``Anomaly detection of {CAN} bus messages through analysis of {ID} sequences,'' in \emph{IEEE Proc. Intelligent Vehicles Symp.}, June 2017.

\bibitem{Stabili2022Consequences}
D.~Stabili, R.~Romagnoli, M.~Marchetti, B.~Sinopoli, and M.~Colajanni, ``Exploring the consequences of cyber attacks on powertrain cyber physical systems,'' 2022.

\end{thebibliography}
